\documentclass[conference,a4paper]{IEEEtran}

\usepackage[utf8]{inputenc}
\usepackage[T1]{fontenc}
\usepackage{url}
\usepackage{hyperref}
\usepackage{ifthen}
\usepackage{cite}
\usepackage[cmex10]{amsmath} %
\usepackage{mathtools}
\usepackage[dvipsnames]{xcolor}
\usepackage{pgfplots}
\usepackage[linesnumbered,ruled]{algorithm2e}
\usepackage{algpseudocode}
\algrenewcommand\algorithmicindent{0.5em}
\usepackage{mathrsfs}
\usepackage{graphicx}
\usepackage{caption}
\usepackage{amsfonts}
\usepackage[normalem]{ulem}
\usepackage{enumitem}
\newtheorem{theorem}{Theorem}
\newtheorem{lemma}{Lemma}
\newtheorem{proposition}{Proposition}

\newtheorem{definition}{Definition}

\renewcommand{\IEEEproofindentspace}{1em}

\newcommand{\F}{\mathbb{F}}

\newcommand{\Fqm}{\mathbb{F}_{q^m}}

\newcommand{\cA}{\mathcal{A}}
\newcommand{\cB}{\mathcal{B}}

\newcommand{\cE}{\mathcal{E}}
\newcommand{\cEbar}{\bar{\mathcal{E}}}

\newcommand{\cS}{\mathcal{S}}

\newcommand{\bcG}{\boldsymbol{\mathcal{G}}}
\newcommand{\bcP}{\boldsymbol{\mathcal{P}}}

\newcommand{\Gspan}{\langle\bcG\rangle}
\newcommand{\bc}{\boldsymbol{c}}
\newcommand{\bw}{\boldsymbol{w}}
\newcommand{\by}{\boldsymbol{y}}
\newcommand{\bz}{\boldsymbol{z}}
\newcommand{\bE}{\boldsymbol{E}}
\newcommand{\bC}{\boldsymbol{C}}
\newcommand{\bM}{\boldsymbol{M}}
\newcommand{\bR}{\boldsymbol{R}}
\newcommand{\List}{\mathcal{L}}

\newcommand{\kGRS}{k_{\mathrm{GRS}}}
\newcommand{\wdeg}{\ensuremath{\mathrm{deg}_{\boldsymbol{w}}}}

\newcommand{\Apara}{\mathrm{A}(\mathcal{S},\mathcal{B},\kGRS)}
\newcommand{\IApara}[1]{\mathcal{I}\mathrm{A}(\mathcal{S},\mathcal{B},\kGRS; #1)}

\newcommand{\cEfree}{\bar{\cE}}
\newcommand{\PsetE}{\bcP'_{\cE}}
\newcommand{\PsetEfree}{\bcP'_{\bar{\cE}}}

\DeclareMathOperator{\colsupp}{colsupp}
\DeclareMathOperator{\resultant}{Resultant}

\renewcommand{\baselinestretch}{0.95389} %

\IEEEoverridecommandlockouts
\begin{document}
\title{List Decoding of 2-Interleaved Binary Alternant Codes}

\author{%
  \IEEEauthorblockN{Chih-Chiang Huang\IEEEauthorrefmark{1}, Hedongliang Liu\IEEEauthorrefmark{1}, Lukas Holzbaur\IEEEauthorrefmark{1}, Sven Puchinger\IEEEauthorrefmark{2}, and Antonia Wachter-Zeh\IEEEauthorrefmark{1}
  }
  \IEEEauthorblockA{\IEEEauthorrefmark{1}Insitute for Communications Engineering,
                    Technical University of Munich, Germany\\
                    }
\IEEEauthorblockA{\IEEEauthorrefmark{2}Hensoldt Sensors GmbH, Ulm, Germany\thanks{The work of C.C.~Huang was supported by the Tsung Cho Chang Foundation. The work of H.~Liu has been supported by a German Israeli Project Cooperation (DIP) grant under grant no.~PE2398/1-1 and KR3517/9-1. The work of L.~Holzbaur and A.~Wachter-Zeh was supported by the German Research Foundation (Deutsche Forschungsgemeinschaft, DFG) under Grant No. WA3907/1-1. S.~Puchinger was supported by the European Union’s Horizon 2020 research and innovation program under the Marie Sklodowska-Curie grant agreement no. 713683 and by the European
Research Council (ERC) under the European Union’s Horizon 2020 research
and innovation programme (grant agreement no. 801434). This work was done while S.~Puchinger was with the Department of Applied Mathematics and Computer Science, Technical
University of Denmark (DTU), Kongens Lyngby, Denmark, and the Department of Electrical and Computer Engineering, Technical University of Munich, Germany.}}
}

\maketitle

\begin{abstract}
   This paper is concerned with list decoding of $2$-interleaved {binary} alternant codes.
  The principle of the proposed algorithm is based on a combination of a list decoding algorithm for (interleaved) Reed-Solomon codes and an algorithm for (non-interleaved) alternant codes. A new upper bound on the decoding radius is derived and the list size is shown to scale polynomially in the code parameters. While it remains an open problem whether this upper bound is achievable, the provided simulation results show that a decoding radius exceeding the binary Johnson radius can be achieved with a high probability of decoding success by the proposed algorithm.
\end{abstract}

\section{Introduction}

Given a received word, a unique decoder returns (at most) one codeword within a specified radius of the received word. In contrast, the goal of a list decoder is to return a list $\List$ containing \emph{all} codewords within Hamming distance at most the decoding radius $\tau$ from the received word. 
A code is said to be $\ell$-list-decodable with radius $\tau$ if the returned list is of size at most $\ell$ for a decoding radius of at most $\tau$.
The most important benefit of a list decoding algorithm is that it commonly allows for increasing the decoding radius compared to unique decoding. In particular, it has been shown \cite{johnson1962new} that any code of length $n$ and minimum distance $d$ can be $\ell$-list-decoded up to the field-size-independent \emph{Johnson radius} $\frac{t}{n} < 1-\sqrt{1-\frac{d}{n}}$, while $\ell$ scales polynomially in~$n$. This result was refined in  \cite{bassalygo1965new} to show that any $q$-ary code can be decoded up to the (strictly larger) $q$-ary Johnson radius\footnote{Note that the $q$-ary Johnson radius approaches the field-size-independent Johnson radius as $q\rightarrow \infty$.} $\frac{t}{n} < \theta_q \big(1-\sqrt{1-\frac{d}{\theta_q n}}\big)$, where $\theta_q = q-\frac{1}{q}$.

Generalized Reed-Solomon (GRS) codes and their subfield subcodes, referred to as \emph{alternant codes}\footnote{This class of codes contains some of the most popular codes over small fields, such as BCH and Goppa codes.}, are among the most popular classes of algebraic codes and their (list-) decoding has attracted considerable attention by researchers.
In 1997, Sudan \cite{Sud97} presented an interpolation-based list decoding algorithm for low-rate GRS codes based on a generalization of the well-known Welch-Berlekamp algorithm. In 1999, Guruswami and Sudan improved the algorithm by introducing the idea of assigning higher multiplicities to points in the interpolation~\cite{782097}. 
At the cost of an increased interpolation complexity, this algorithm allows for list-decoding GRS codes up to the field-size-independent Johnson bound.
In 2003, Koetter and Vardy further improved the algorithm by introducing varying multiplicities \cite{algebraic03koetter} in order to use soft-information on the reliability of the received symbols. When applying this algorithm to alternant codes, the fact that the symbols in both the codewords and the received word are contained in a subfield can be regarded as soft information. In \cite{6089384} it was shown that choosing the multiplicities accordingly further increases the decoding radius to the binary ($q=2$) Johnson radius. 

Interleaving is another powerful method to achieve a larger decoding radius. In interleaved decoding, each codeword is a matrix where every row is a codeword of a given code and the weight of the error is determined by the number of non-zero columns in the error matrix. This concept has been applied to GRS codes \cite{krachkovsky1997decoding, schmidt2009collaborative, coppersmith2003reconstructing, Par07, AWDecIRS14}, alternant codes \cite{holzbaur2021alternant}, and algebraic-geometry codes \cite{brown2005improved,puchinger2019improved}.

Parvaresh \cite{Par07} combined list and interleaved decoding by adapting the Guruswami-Sudan (GS) algorithm to the decoding of $2$-interleaved GRS codes. To this end, trivariate polynomials are used to set up the interpolation constraints and the resultants of polynomials are used to recover the codeword. By combining the approaches of interleaved decoding and the GS algorithm, this decoder achieves a larger decoding radius than the GS algorithm, however, at the cost of a small probability of failure. %

In this paper, we propose a list decoding algorithm for $2$-interleaved binary alternant codes that combines the ideas of applying the Koetter-Vardy algorithm \cite{algebraic03koetter} to alternant codes \cite{6089384} and Parvaresh's algorithm to interleaved GRS codes \cite{Par07}. Similar to Parvareh's algorithm, it is difficult to make a precise statement on the decoding radius of this code. Instead, we present an upper bound on this radius, along with simulation results showing that the decoding radius of the algorithm exceeds the decoding radii of all other algorithms known in literature for the chosen parameters. The drawback of the presented algorithm is that decoding is not guaranteed to succeed (similar to \cite{Par07}). However, the simulation results indicate that this probability of failure is small, if the parameters of the algorithm are chosen suitably.

Due to space limitations, some proofs are omitted and can be found in the extended version \cite{full_version} of this work.
\section{Preliminaries}

Let $[n]$ be the set of integers $\{1,\dots, n\}$. For a prime power $q$ and an integer $m$, denote by $\F_q$ the finite field with $q$ elements and by $\Fqm$ its extension field. Let $\F^*_q \coloneqq \F_q\setminus \{0\}$ and $\mathbb{N}_0$ be the natural numbers including zero. For two sets $\cA,\cB$, denote their Cartesian product by $\cA\times\cB \coloneqq \{(a,b) \ |\ a\in \cA, b\in \cB\}$.
Denote a linear code of length $n$ and dimension $k$ over the field $\F_q$ by $[n,k]_q$. For a matrix $\bE\in \F_q^{M\times n}$ define $\colsupp(\bE)\subseteq [n]$ as the set of indices of the non-zero columns in $\bE$.
 Let $\F_q[X,Y,Z]$ be the ring of polynomials in $X,Y,$ and $Z$ with coefficients in $\F_q$. 
If clear from context, we omit the variables from the notation and simply write $G$ for a polynomial $G(X,Y,Z) \in \F_q[X,Y,Z]$.

\begin{definition}[Weighted Degree of Trivariate Monomials] \label{def:weighteddegree}
Let $\boldsymbol{w}=\left ( w_{1},w_{2},w_{3} \right )$ be a tuple  of positive integers. The {($\boldsymbol{w}$-)} \!weighted degree of a trivariate monomial $ X^{a}Y^{b}Z^{c}$, is defined as
\begin{equation*}
	\mathrm{deg}_{\boldsymbol{w}} ( X^{a}Y^{b}Z^{c} )\coloneqq w_{1}a+w_{2}b+w_{3}c\ .
\end{equation*}
\end{definition}

We define the $X$-degree (resp.~$Y,Z$) of a polynomial $G(X,Y,Z) \in \mathbb{F}_q[X,Y,Z]$ to be $\deg_X(G) \coloneqq \deg_{(1,0,0)}(G)$.
\begin{definition}[Monomial Ordering]\label{def:ordering}
For two monomials $X^aY^bZ^c$ and $X^uY^vZ^s$, define a monomial ordering $\prec_{\bw}$ with $X^aY^bZ^c\prec_{\bw}X^uY^vZ^s$ if $\wdeg (X^aY^bZ^c)<\wdeg (X^uY^vZ^s)$. If the weighted degrees are equal, then $X^aY^bZ^c\prec_{\bw}X^uY^vZ^s$ if and only if $a<u$ or $a=u$ and $b<v$. The \emph{leading monomial} of a polynomial is the largest term under the ordering $\prec_{\bw}$. The weighted degree of a polynomial $G(X,Y,Z) \in \mathbb{F}_q[X,Y,Z]$ is the weighted degree of its leading monomial.
\end{definition}

Note that any polynomials $Q\left ( X,Y,Z \right ),P\left ( X,Y,Z \right ) \in \F_{2^m}[X,Y,Z]$ can be written in the form
\begin{equation*}%
	\begin{aligned}
	    &\scalebox{0.8}{$Q\left ( X,Y,Z \right )=\mathrm{Coeff}_{Q}\left(X,Y\right)\left ( Z-q_{1}\left ( X,Y \right ) \right )\left ( Z-q_{2}\left ( X,Y \right ) \right )\ldots,$} \\
	    &\scalebox{0.8}{$P\left ( X,Y,Z \right )=\mathrm{Coeff}_{P}\left(X,Y\right)\left ( Z-p_{1}\left ( X,Y \right ) \right )\left ( Z-p_{2}\left ( X,Y \right ) \right )\ldots$},
	\end{aligned}
\end{equation*}
where the coefficients $\mathrm{Coeff}_{Q}\left(X,Y\right)$ and $\mathrm{Coeff}_{P}\left(X,Y\right)$ are functions in $X$ and $Y$ and $q_{i}\left ( X,Y \right ),\ p_{j}\left ( X,Y \right )$ are the $Z$-roots of $Q$ and $P$, respectively.

\begin{definition}[Resultant{\cite{macaulay1902some,andrea2001explicit}}]\label{def:Resultant}
The resultant of two polynomials $Q\left ( X,Y,Z \right )$ and $P\left ( X,Y,Z \right )$ w.r.t.~$Z$ is
\begin{equation*}\label{eq:IG2CFactorQP}
	\begin{aligned}
	    H_{Z}\left(X,Y\right) &= \resultant\left ( Q\left ( X,Y,Z \right ),P\left ( X,Y,Z \right );Z \right ) \\
	    &\coloneqq \mathrm{Coeff}\left(X,Y\right)\prod_{i,j}\left ( q_{i}\left ( X,Y \right ) - p_{j}\left ( X,Y \right ) \right ),
	\end{aligned}
\end{equation*}
where $\mathrm{Coeff}\left(X,Y\right) = \mathrm{Coeff}_{Q}^{\deg_Z\left(P\right)} \mathrm{Coeff}_{P}^{\deg_Z\left ( Q\right ) }$.
\end{definition}

\begin{definition}[Generalized Reed-Solomon Code]\label{def:GRS}
    Denote by $\cS=\left \{\alpha_{1},\alpha_{2},\ldots,\alpha_{n}\right \} \subseteq \Fqm$ a set of $n$ distinct code locators and by $\mathcal{B}=\left \{\beta_{1},\beta_{2},\ldots,\beta_{n}\right \}\subseteq \Fqm^*$ a set of column multipliers.
    An $[ n,k_{\mathrm{GRS}}]_{q^m}$ Generalized Reed-Solomon (GRS) is defined by
       \begin{align*}
         \mathrm{GRS}_{q^m}( \cS,\mathcal{B},k_{\mathrm{GRS}}& )\coloneqq \{ \big(\beta_1 f(\alpha_1), \ldots , \beta_n f(\alpha_n)\big)  \\
          &  | \ f ( X )\in\Fqm [ X  ],\ \deg \big( f( X ) \big) < k_{\mathrm{GRS}}  \}.
    \end{align*}
\end{definition}

It is well-knonw that GRS codes are maximum distance separable (MDS) codes, i.e., of minimum distance $d=n-k_{\mathrm{GRS}}+1$.

\begin{definition}[Binary Alternant Code]\label{def:alternantCode}
  Let $\mathrm{GRS}_{2^m}( \cS,\mathcal{B},k_{\mathrm{GRS}})$ be as in Definition~\ref{def:GRS}. Its \emph{alternant code} is defined as
  \begin{align*}
    \mathrm{A}( \cS,\mathcal{B},k_{\mathrm{GRS}}) \coloneqq \mathrm{GRS}_{2^m}( \cS,\mathcal{B},k_{\mathrm{GRS}}) \cap \F_2^{n} .
  \end{align*}
\end{definition}
Note that $\kGRS$ is not the dimension of the alternant code, but only serves as a (loose) upper bound. However, as the parameters $\cS,\mathcal{B}$, and $k_{\mathrm{GRS}}$ uniquely specify a GRS code, they also determine the corresponding alternant code.

\begin{definition}[$2$-Interleaved Alternant Code]
    Let $\mathrm{A}( \cS,\mathcal{B},k_{\mathrm{GRS}})$ be as in Definition~\ref{def:alternantCode}. Define the $2$-interleaved alternant code as
  \begin{align*}
    \IApara{2} \!\coloneqq\! \left\{
    \begin{pmatrix}
      \bc^{(1)} \\
      \bc^{(2)} \\
    \end{pmatrix}
    \Bigg| 
    \begin{aligned}
        &\bc^{(1)}, \bc^{(2)} \in \mathrm{A}( \cS,\mathcal{B},k_{\mathrm{GRS}})\\
    \end{aligned}\right\}.
  \end{align*}
\end{definition}

Throughout this paper, we associate the codewords $\bc^{(1)},\bc^{(2)} \in \Apara$ with their respective evaluation polynomials $f(X),g(X) \in \F_{2^m}[X]$ by
\begin{align*}
	\bc^{\left ( 1 \right )}&=\big(\beta_1 f(\alpha_1), \ldots , \beta_n f(\alpha_n)\big),\\
	\bc^{\left ( 2 \right )}&=\big(\beta_1 g(\alpha_1), \ldots , \beta_n g(\alpha_n)\big).
\end{align*}

Let $\boldsymbol{R}\in \F_2^{2\times n}$ be a received word with 
\begin{equation}\label{eq:IG2CReceivedWordR}
	\boldsymbol{R} = 
	\begin{pmatrix}
	\by\\
	\bz 
	\end{pmatrix} = 
	\begin{pmatrix}
		\boldsymbol{c}^{\left ( 1 \right )}\\ 
		\boldsymbol{c}^{\left ( 2 \right )}\\ 
	\end{pmatrix}  
	+
	\begin{pmatrix}
		\boldsymbol{e}^{\left ( 1 \right )}\\ 
		\boldsymbol{e}^{\left ( 2 \right )}\\ 
	\end{pmatrix} = 
	\boldsymbol{C} + \boldsymbol{E}
	,
\end{equation}
where $\boldsymbol{C} \in \IApara{2}$.
The \emph{error positions} of $\bE$ are given by $\cE\coloneqq\colsupp(\bE)$. Denote the \emph{error-free positions} by $\cEbar\coloneqq [n]\setminus \cE$.

Define the set $\boldsymbol{\mathcal{P}}$ of $n$ received points as
\begin{equation*}%
	\begin{aligned}
		\boldsymbol{\mathcal{P}}\coloneqq \left \{ \left ( x_{1},y_{1},z_{1} \right ),
		\ldots,\left ( x_{n},y_{n},z_{n} \right ) \right \} \subset \cS\times \F_2\times \F_2,
	\end{aligned}
\end{equation*}
where $\left ( x_{s},y_{s},z_{s} \right ) = ( \alpha _{s},\beta _{s}f\left ( \alpha _{s} \right ) +e_{s}^{\left ( 1 \right )},\beta _{s}g\left ( \alpha _{s} \right ) +e_{s}^{\left ( 2 \right )} ),$ $\forall s \in [n]$
and a modified set $\boldsymbol{\mathcal{P}}'$ of received points as
\begin{equation*}
	\begin{aligned}
		\boldsymbol{\mathcal{P}}'\coloneqq 
		\{ \left ( x_{1},\beta_{1}^{-1}y_{1},\beta_{1}^{-1}z_{1} \right ), 
		\ldots,\left ( x_{n},\beta_{n}^{-1}y_{n},\beta_{n}^{-1}z_{n} \right ) \}.
	\end{aligned}
\end{equation*}
Denote by $\bcP'_{\cE}$ and $\PsetEfree$ the subsets of $\bcP'$ corresponding to the erroneous positions in $\cE$ and the error-free positions in $\cEfree$, respectively.

\begin{definition}[Hasse Derivatives of Trivariate Polynomials {\cite{Par07}}] \label{def:IG2C3DHasseDerivative}
For a polynomial $G \left ( X,Y,Z \right )\in\mathbb{F}_{q}\left [ X,Y,Z \right ]$, the $\left ( a,b,c \right )$-th Hasse derivative of $G \left ( X,Y,Z \right )$, denoted by $\mathscr{D}_{a,b,c}\left [ G \left ( X,Y,Z \right ) \right ]$, is defined as
\begin{equation*}
\begin{aligned}
	\mathscr{D}&_{a,b,c}\left [ G \left ( X,Y,Z \right ) \right ]\coloneqq\\
	&\sum_{p=a}^{\infty}\sum_{q=b}^{\infty}\sum_{r=c}^{\infty}\binom{p}{a}\binom{q}{b}\binom{r}{c}\  g_{p,q,r} \  X^{p-a}Y^{q-b}Z^{r-c},
\end{aligned}
\end{equation*}
where $g_{p,q,r}$ denotes the coefficient of the term $X^{p}Y^{q}Z^{r}$ in $G \left ( X,Y,Z \right )$.
\end{definition}

\begin{definition}[Multiplicity of Trivariate Polynomials {\cite[Def.~2.1]{Par07}}] \label{def:IG2C3DMultiplicity}
Let $P=\left ( x_{s},y_{s},z_{s} \right ) \in \mathbb{F}_{q}\times \mathbb{F}_{q}\times \mathbb{F}_{q}$. The interpolation polynomial $G \left ( X,Y,Z \right )$ is said to pass through the point $P=\left ( x_{s},y_{s},z_{s} \right )$ with multiplicity $m_{1}$, or to have a zero of multiplicity $m_{1}$ at the point, if
\begin{equation*}
	\left. \begin{matrix}
		\mathscr{D}_{a,b,c}\left [ G \left ( X,Y,Z \right ) \right ]
	\end{matrix}\right|_{P}=0, \  \forall \  a,b,c\in\mathbb{N}_{0},\ a+b+c<m_{1}.
\end{equation*}
\end{definition}
By this definition, the interpolation constraint of passing through a point with multiplicity $m_1$ imposes $\binom{m_1+2}{3}$ linear constraints on the coefficients of $G(X,Y,Z)$.

In this work, we fix the weight $\boldsymbol{w}$ of the $\boldsymbol{w}$-weighted degree to be $\boldsymbol{w}=\left ( 1,k_{\mathrm{GRS}}-1,k_{\mathrm{GRS}}-1 \right )$, where $\kGRS$ denotes the dimension of the GRS containing the alternant code (see Definition~\ref{def:alternantCode}) under consideration.

\section{List Decoding Algorithm for $2$-Interleaved Binary Alternant Codes}
\label{sec:algorithm}
\subsection{Sketch of the Algorithm}
\label{sec:sketchAlgorithm}
Let $\boldsymbol{R}$ be a received word as defined in (\ref{eq:IG2CReceivedWordR}) and $m_1, m_2$ be integers such that $0\leq m_2<m_1$. 
We give a sketch of the proposed decoding algorithm consisting of the following steps: initialization, interpolation, and recovery.

\begin{enumerate}[leftmargin=*]
	\item \emph{Initialization:} Let
	\begin{equation}\label{eq:IG2CDelta}
	\begin{split}
	   \Delta &\coloneqq (n\left ( k_{\mathrm{GRS}}-1 \right )^{2}m_{1}\left ( m_{1}+1 \right )\left ( m_{1}+2 \right )\\
	   &\qquad +3n\left ( k_{\mathrm{GRS}}-1 \right )^{2}m_{2}\left ( m_{2}+1 \right )\left ( m_{2}+2 \right ))^{\frac{1}{3}}
	\end{split}
    \end{equation}
    and $\mu \coloneqq \left \lceil \frac{\Delta}{k_{\mathrm{GRS}}-1} \right \rceil$.
	Initialize a set $\boldsymbol{\mathcal{G}}^{(0)}$ of trivariate polynomials by
\begin{align*} %
	    \boldsymbol{\mathcal{G}}^{\left ( 0 \right )}&\coloneqq \left \{ G_{1}^{\left ( 0 \right )},G_{2}^{\left ( 0 \right )},\ldots ,G_{l}^{\left ( 0 \right )} \right \}
	    \\
	    &= \left \{ Y^{\alpha}Z^{\beta}:\forall \alpha,\beta\in \mathbb{N}_{0} \  \mathrm{s.t.}\   \alpha+\beta<\mu \right \}.
\end{align*}
Note that the size of $\bcG^{(0)}$ is $l\coloneqq \frac{\mu \left ( \mu+1 \right )}{2}$. %
	
	\item \emph{Interpolation:}\label{step:Interpolation}
	Find a Groebner basis\footnote{For the full algorithm to compute such a Groebner basis, see the extended version \cite[Algorithm~2]{full_version}, which is adapted from the algorithm given in \cite[Section 2.4]{Par07}.} $\bcG$ of the ideal of all polynomials that, for all $s\in [n]$, satisfy the interpolation constraints:
	\begin{itemize}
	    \item passes through $  ( \alpha _{s},\beta_{s}^{-1}y_{s}, \beta_{s}^{-1}z_{s}   )\in\bcP'$ with multiplicity $m_{1}$,
	    \item passes through $  ( \alpha _{s},\beta_{s}^{-1}\gamma_{1}, \beta_{s}^{-1}\gamma_{2}   )$ with multiplicity $m_{2}$, $\forall (\gamma_{1},\gamma_{2}) \in \mathbb{F}_{2}\times\mathbb{F}_{2} \setminus    \{  ( y_{s},z_{s}   )  \} $.
	\end{itemize}

	\item \emph{Recovery:} \label{step:Recovery} 
Find the pair of interpolation polynomials $G_r,G_s\in\bcG$ of the lowest weighted-degree that do not have a common factor that is a polynomial in $Y$ or $Z$. If no such pair of $G_r,G_s$ can be found in $\bcG$, return a \texttt{decoding failure}. Otherwise, denote $\widehat{\Delta}=\max\{\wdeg(G_r),\wdeg(G_s)\}$. 
Factorize the resultants (see Definition~\ref{def:Resultant}) of $G_r$ and $G_s$ with respect to $Y$ and $Z$ to get all the factors $Y-\hat{f}(X)$ and $Z-\hat{g}(X)$. 
Reconstruct the codewords $\hat{\bc}^{(1)}$ and $\hat{\bc}^{(2)}$ by evaluating $\hat{f}(X)$ and $\hat{g}(X)$ respectively. 
Append all the interleaved codewords to the returned list $\hat{\List}$. 
Return the \emph{achieved} decoding radius $\hat{\tau}\coloneqq\frac{n-\widehat{\Delta}/{m_{1}}}{1-{m_{2}}/{m_{1}}}$ and the list $\hat{\List}$ containing all codewords within radius $\hat{\tau}$ of the received word. 
\end{enumerate}
For details, please see the full algorithm in the extended version of this paper \cite[Appendix]{full_version}.

\subsection{Upper Bound on Decoding Radius of the Algorithm}
Let $\bcG$ be the Groebner basis for the ideal of all polynomials fulfilling the interpolation constraints in Section~\ref{sec:sketchAlgorithm}.
The following lemma gives a condition for the existence of a polynomial in $\bcG$ of low weighted degree that fulfills the interpolation constraints.
\begin{lemma}\label{lemma:IG2CUpperBoundOfWeightedDegree}
For $\Delta$ as in~\eqref{eq:IG2CDelta},
there always exists a nonzero $G\left ( X,Y,Z \right )\in\bcG$ with
\begin{equation*}
	\begin{aligned}
		\wdeg(G\left ( X,Y,Z \right )) \leq  \Delta.
	\end{aligned}
\end{equation*}
\end{lemma}
\begin{IEEEproof}
By Definition \ref{def:IG2C3DMultiplicity}, there are
$\binom{m_1+2}{3}$
linear constraints imposed on a polynomial to pass through a point with multiplicity $m_{1}$. Therefore, the total number of linear constraints imposed on the coefficients of any polynomial fulfilling the interpolation constraints is
\begin{equation*} \label{eq:IG2CInterpolationCost}
	\begin{aligned}
		C\left ( m_{1},m_{2},n \right ) \coloneqq n\cdot \binom{m_1+2}{3}+3n\cdot \binom{m_2+2}{3}\ .
	\end{aligned}
\end{equation*}
 We call $C\left ( m_{1},m_{2},n \right )$ the interpolation cost.
 Denote by $N(\Delta)$ be the number of trivariate monomials whose weighted degree is at most $\Delta$. From the proof of~\cite[Lemma 2.1]{Par07}, $N(\Delta)>\frac{\Delta^{3}}{6 \left( k_{\mathrm{GRS}-1} \right )^{2} }$. Finding a nonzero polynomial in $\F_{2^m}[X,Y,Z]$ that fulfills the interpolation constraints of Section~\ref{sec:sketchAlgorithm}, \ref{step:Interpolation}) is equivalent to solving a linear system of equations with at most $C(m_1,m_2,n)$ constraints for at least $N(\Delta)$ unknowns. A non-zero solution is guaranteed to exist if $N(\Delta)>C(m_1,m_2,n)$. This inequality always holds with $\Delta$ as in~\eqref{eq:IG2CDelta}. Since $\bcG$ is a Groebner basis, it contains the polynomial of lowest weighted degree that fulfills the constraints and the lemma statement follows.
\end{IEEEproof}
\par
Denote by $\langle\bcG\rangle$ the set of polynomials spanned by $\bcG$.
The following lemma gives a condition on the weighted degree of the interpolation polynomial $G(X,Y,Z)\in\langle\bcG\rangle$ such that $f(X)$ and $g(X)$ are a $Y$-root and $Z$-root of $G(X,Y,Z)$, respectively.
\begin{lemma}\label{lemma:IG2CGEquiv0}
For a received word with $t$ errors, let $\bcG$ be the Groebner basis as in Section~\ref{sec:sketchAlgorithm}. For any nonzero $G(X,Y,Z)\in\Gspan$ with
\begin{equation*}%
\mathrm{deg}_{\boldsymbol{w}}\left(G\left ( X,Y,Z \right )\right)<m_{1}\left(n-t\right)+m_{2}t\ ,
\end{equation*}
it holds that $G\left ( X,f\left ( X \right ),g\left ( X \right ) \right ) \equiv 0$.
\end{lemma}
\begin{IEEEproof}
Let $G(X,Y,Z)$ be a nonzero polynomial in $\Gspan$ and $G(X, f(X), g(X))=0$.
Recall from Definition \ref{def:IG2C3DHasseDerivative} and \ref{def:IG2C3DMultiplicity} that for any point $(x_s, y_s, z_s)$ that $G(X,Y,Z)$ passes through, we have
\begin{equation} \label{eq:IG2CCoefficientCompare}
	\begin{aligned}
		&G \left ( X+x_{s},Y+y_{s},Z+z_{s} \right ) \\
		= &\sum_{i} \sum_{j} \sum_{k}\ g_{i,j,k}\ \left ( X+x_{s} \right )^{i}\left ( Y+y_{s} \right )^{j}\left ( Z+z_{s} \right )^{k} \\
		= &\sum_{i} \sum_{j} \sum_{k}\ \left( \scalebox{0.78}{$\sum_{p}\sum_{q}\sum_{r}\binom{p}{i}\binom{q}{j}\binom{r}{k}\  g_{p,q,r} \  x_{s}^{p-i}y_{s}^{q-j}z_{s}^{r-k}$} \right ) \cdot X^{i} Y^{j} Z^{k} \\
		= &\sum_{i} \sum_{j} \sum_{k}\
			\mathscr{D}_{i,j,k}\left [ G \left ( X,Y,Z \right ) \right ]
		|_{\left( x_{s},y_{s},z_{s} \right)}
		\cdot X ^{i} Y ^{j} Z ^{k}.
	\end{aligned}
\end{equation}
Since $G(X,Y,Z)\in\langle\bcG\rangle$, $G \left ( X,Y,Z \right )$ passes through the $\left(n-t\right)$ error-free received points in $\PsetEfree$ with multiplicity $m_{1}$. Therefore, by Definition \ref{def:IG2C3DMultiplicity}, we have
\begin{equation} \label{eq:IG2CGoppaTrivarriateZero}
    \begin{aligned}
    	\mathscr{D}_{i,j,k}&\left [ G \left ( X,Y,Z \right ) \right ]|_{\left ( \alpha_{s},f\left(\alpha_{s}\right),g\left(\alpha_{s}\right) \right )}=0, \\
    	& \forall i,j,k\in\mathbb{N}_{0}\ \text{with}\ i+j+k<m_{1}, \forall s \in \cEfree .
    \end{aligned}
\end{equation}
Hence, for the error-free points $(x_{s},y_s,z_s) \in \PsetEfree$, \eqref{eq:IG2CCoefficientCompare} becomes
\begin{equation*}
	\begin{aligned}
	G \left ( X+x_{s},Y+y_{s},Z+z_{s} \right )
	= \underset{i+j+k\geq m_{1}}{{\sum_{i} \sum_{j} \sum_{k}}}\ g'_{i,j,k} \  X ^{i} Y ^{j} Z ^{k},
	\end{aligned}
\end{equation*}
for some $g'_{i,j,k}\in \F_{2^m}$. %
Let
\begin{equation*}
	\begin{aligned}
		P\left(X\right)
		\coloneqq &\ G\left ( X,f\left ( X \right ),g\left ( X \right ) \right ) \\
		= &\ \scalebox{0.78}{$\underset{i+j+k>=m_{1}}{{\sum_{i} \sum_{j} \sum_{k}}}\ g'_{i,j,k} \  \left( X-\alpha_{s} \right) ^{i}
		\left( f \left(  X\right)-f\left( \alpha_{s} \right) \right) ^{j}
		\left( g \left(  X\right)-g\left( \alpha_{s} \right) \right) ^{k}$}.
	\end{aligned}
\end{equation*}
Notice that
\begin{align*}
    \left. \left( f \left(  X\right)-f\left( \alpha_{s} \right) \right)\right |_{X=\alpha_{s}} &= 0 \\
    \left. \left( g \left(  X\right)-g\left( \alpha_{s} \right) \right)\right |_{X=\alpha_{s}} &= 0 \ ,
\end{align*}
so $\left( X-\alpha_{s} \right)$ divides $\left( f \left(  X\right)-f\left( \alpha_{s} \right) \right)$ and $\left( g \left(  X\right)-g\left( \alpha_{s} \right) \right)$. It follows that
\begin{equation*}
	\left( X-\alpha_{s} \right) ^{m_{1}} | P\left(X\right),\ \forall s \in \cEfree \ .
\end{equation*}

Therefore, $P(X)$ vanishes at the $(n-t)$ error-free points $(\alpha_s, f(\alpha_s), g(\alpha_s))\in\PsetEfree$ with multiplicity $m_1$.

Similarly, as $G( X,Y,Z)$ also passes through the erroneous points in $\PsetE$
with multiplicity $m_{2}$, $P(X)$ vanishes at the $t$ erroneous points $( \alpha_{s},f ( \alpha_{s} ), g(\alpha_s))\in \PsetE $ with multiplicity $m_{2}$. %

Therefore, since $G\neq 0$ by definition,
\begin{equation*}
	\deg P(X)
	\geq m_{1}\left ( n-t \right )+m_{2}t \ .
\end{equation*}
Recall that we fix the weights of $\wdeg$ to be $\bw=( 1,k_{\mathrm{GRS}}-1,k_{\mathrm{GRS}}-1)$. Since $f\left ( X \right )$ and $g\left ( X \right )$ have degree at most $k_{\mathrm{GRS}}-1$, this implies
\begin{equation*}
	\wdeg\left(G\left ( X,Y,Z \right )\right)\geq \mathrm{deg}_{X}\left(G\left ( X,f\left ( X \right ),g\left ( X \right ) \right )\right) \ .
\end{equation*}
Hence, if $\wdeg\left(G\left ( X,Y,Z \right )\right)<m\left(n-t\right)+m_{2}t$, it holds that
\begin{equation*}
        G\left ( X,f\left ( X \right ),g\left ( X \right ) \right ) \equiv 0 \ .%
\end{equation*}
\end{IEEEproof}

Using this result, we now present an upper bound on the number of errors such that there exists at least one interpolation polynomial in $\bcG$ fulfilling Lemma \ref{lemma:IG2CGEquiv0}.
\begin{theorem}\label{lemma:IG2CDecodingRadius1Polynomial}
Let $\Delta, m_1, m_2,$ and $\bcG$ be as in Section~\ref{sec:sketchAlgorithm}.
Then, for any received word with $t$ errors, where
\begin{equation}\label{eq:radiusUpperBound}
t\leq \frac{n-{\Delta}/{m_{1}}}{1-{m_{2}}/{m_{1}}},
\end{equation}
there exists a $G(X,Y,Z)\in \bcG$ such that
$G\left ( X,f\left ( X \right ),g\left ( X \right ) \right ) = 0.$
\end{theorem}
\begin{IEEEproof}
The upper bound follows from Lemma~\ref{lemma:IG2CUpperBoundOfWeightedDegree} and Lemma~\ref{lemma:IG2CGEquiv0} by setting $\Delta\leq m_{1}\left(n-t\right)+m_{2}t$ and solving the inequality for $t$.
\end{IEEEproof}

Note that the recovery step in Section~\ref{sec:sketchAlgorithm}, \ref{step:Recovery}) requires at least two polynomials
in $\bcG$ to have $f(X)$ and $g(X)$ as their $Y$- and $Z$-roots. A polynomial in $\bcG$ is guaranteed to have this property, if its weighted degree fulfills the restriction of Lemma~\ref{lemma:IG2CGEquiv0}. The following theorem gives a bound on the number of errors such that the existence of at least two polynomials in $\bcG$ of sufficiently small weighted degree is guaranteed.

\begin{theorem}\label{theorem:IG2CDecodingRadiusUpperBound}
Consider a codeword of an $\IApara{2}$ code corrupted by $t$ errors. Denote $n=|\cS|$ and $d = n-\kGRS+1$.
Let $m_1, m_2$ and $\bcG$ be as in Section~\ref{sec:sketchAlgorithm}.
Let
\begin{align}\label{eq:normalized_radius}
    \sigma\left(\frac{m_2}{m_1}\right) \coloneqq  \frac{1-(1-d/n)(1+\nu)/2}{1-m_2/m_1},
\end{align}
where $\nu=\sqrt{
        \frac{4}{3\left (1-{d}/{n}  \right )}-\frac{1}{3}+\frac{4}{\left (1-{d}/{n}  \right )} (\frac{m_{2}}{m_{1}})^{3} }.$
If the number of errors is
\begin{equation}\label{eq:upperBoundOverRatioOfm2m1}
	\begin{aligned}
	    t< n \cdot \sigma \left ( \frac{m_{2}}{m_{1}} \right ),
	\end{aligned}
\end{equation}
then there exist at least two polynomials $Q(X,Y,Z),P(X,Y,Z) \in \bcG$
such that
\begin{equation*} %
	\begin{aligned}
	    &Q\left ( X,f\left ( X \right ),g\left ( X \right ) \right ) = 0 \text{ and }
	    P\left ( X,f\left ( X \right ),g\left ( X \right ) \right ) = 0.
	\end{aligned}
\end{equation*}
\end{theorem}
\begin{IEEEproof}
Let $G_1( X,Y,Z )\in\bcG$ be the polynomial of smallest weighted degree in $\bcG$ and $\Delta_1=\wdeg ( G_1( X,Y,Z ))$. Note that $G_1$ is unique by definition of the monomial ordering $\prec_{\bw}$ (see Definition~\ref{def:ordering}).

We first show that $G_1\neq 0$ only if
\begin{align}\label{eq:lb_Delta1}
    \Delta_1 > m_1(\kGRS-1).
\end{align}
Let the leading monomial of $G_{1}( X,Y,Z )$ be $g_{1,A,B,C}X^{A}Y^{B}Z^{C}$, where $A\in \left [0,\Delta_{1}\right]$ and $B, C \in [0,\lfloor \frac{\Delta_{1}}{k_{\mathrm{GRS}}-1}  \rfloor ]$. 
By Definition~\ref{def:weighteddegree}, 
\begin{equation*}
    \Delta_1 = \wdeg G_{1}\left ( X,Y,Z \right )=A+\left(k_{\mathrm{GRS}}-1\right)\left(B+C\right). 
\end{equation*}
By Definition~\ref{def:IG2C3DMultiplicity} and the interpolation constraints, we have 
\begin{equation*}
	\begin{aligned}
    	\mathscr{D}_{a,b,c}&\left [ G_{1} \left ( X,Y,Z \right ) \right ]|_{\left ( \alpha _{s},\beta_{s}^{-1}y_{s}, \beta_{s}^{-1}z_{s} \right )}=0, \\ &\forall \  a,b,c\in\mathbb{N}_{0}\ \text{with}\ a+b+c<m_{1}, \forall s \in [n].
	\end{aligned}
\end{equation*}

Denote $A'\coloneqq \mathrm{min}\left \{ m_{1}-B-C-1,A \right \}$. 
For $B+C<m_{1}$ and $a\in [0,A']$, since $a+B+C<m_1$, we have
\begin{equation*}
	\begin{aligned}
    	&\left. \begin{matrix}
    		\mathscr{D}_{a,B,C}\left [ G_{1} \left ( X,Y,Z \right ) \right ]
    	\end{matrix}\right|_{\left ( \alpha _{s},\beta_{s}^{-1}y_{s}, \beta_{s}^{-1}z_{s} \right )} \\
    	&\qquad =\sum_{p=a}^{A}\binom{p}{a}g_{1,p,B,C}\ \alpha_{s}^{p-a}\left ( \beta_{s}^{-1}y_{s} \right )^{B-B}\left ( \beta_{s}^{-1}z_{s} \right )^{C-C} \\
    	&\qquad =\sum_{p=a}^{A}\binom{p}{a}g_{1,p,B,C}\ \alpha_{s}^{p-a} = 0 \ .
	\end{aligned}
\end{equation*}
This is a linear system of equations
\begin{equation}\label{eq:IG2CAllZeroG}
	\begin{aligned}
	    &\scalebox{0.8}{$\binom{0}{0}g_{1,0,B,C}\ \alpha_{s}^{0} + \binom{1}{0}g_{1,1,B,C}\ \alpha_{s}^{1} + \ldots + \binom{A}{0}g_{1,A,B,C}\ \alpha_{s}^{A}=0, \ \ \forall \ \alpha_{s}\in\mathcal{L}$}\\
        &\scalebox{0.8}{$\binom{1}{1}g_{1,1,B,C}\ \alpha_{s}^{0} + \binom{2}{1}g_{1,2,B,C}\ \alpha_{s}^{1} + \ldots + \binom{A}{1}g_{1,A,B,C}\ \alpha_{s}^{A-1}=0, \ \ \forall \ \alpha_{s}\in\mathcal{L}$}\\
        &\ \ \ \ \ \ \ \ \ \ \ \ \ \vdots \\
        &\scalebox{0.7}{$\binom{A'}{A'}g_{1,A',B,C}\ \alpha_{s}^{0} + \binom{A'+1}{A'}g_{1,A'+1,B,C}\ \alpha_{s}^{1} + \ldots + \binom{A}{A'}g_{1,A,B,C}\ \alpha_{s}^{A-A'}=0, \ \ \forall \ \alpha_{s}\in\mathcal{L}.$}
    \end{aligned}
\end{equation}
with $n\left(A'+1\right)$ linear constraints and $A+1$ unknowns $g_{1,a,B,C}, a\in[0,A]$.
If there are more linear constraints than unknowns in~(\ref{eq:IG2CAllZeroG}), then the only solution is $g_{1,i,B,C}=0,\ \forall \ i\in[0,A]$. 
Therefore, we require $n(A'+1)\leq A+1 $ to obtain a nonzero solution.

Recall that $A'=\mathrm{min}\left \{ m_{1}-B-C-1,A \right \}$. For the case $A<m_1-B-C-1$, we always have $$n\left(A'+1\right) =n\left(A+1\right)>A+1,$$ as this inequality holds for any non-trivial code with $n>1$. 
For the other case, assume that \eqref{eq:lb_Delta1} does not hold, i.e.,
\begin{align*}
\Delta_1 = A+\left(k_{\mathrm{GRS}}-1\right)\left(B+C\right)<m_{1}\left(k_{\mathrm{GRS}}-1\right),
\end{align*}
which leads to $$A<\left(k_{\mathrm{GRS}}-1\right)\left(m_{1}-B-C\right).$$ %
Then,
\begin{align*}
    n\left(A'+1\right) &= n\left(m_{1}-B-C\right)\\
    &>\left(k_{\mathrm{GRS}}-1\right)\left(m_{1}-B-C\right)>A.
\end{align*}
Therefore, \eqref{eq:lb_Delta1} is a necessary condition such that there is a nonzero solution to~\eqref{eq:IG2CAllZeroG}, i.e., a nonzero $G_1$ of smallest weighted degree that fulfills the interpolation constraints.

Let $G_2( X,Y,Z )\in\bcG$ be the polynomial of second smallest weighted degree in $\bcG$ and $\Delta_2=\wdeg G_2( X,Y,Z )$. Note that $G_1\prec_{\bw} G_2$ and therefore $\Delta_2\geq \Delta_1$ according to Definition~\ref{def:ordering}.
By \cite[Proof of Lemma~2.7]{Par07}, the number of monomials in $G_2$ of weighted degree smaller than $\Delta_2$ is at most $$\frac{\Delta_{2}^{3}}{6\left ( k_{\mathrm{GRS}}-1 \right )^{2}}-\frac{\left (\Delta_{2}-\Delta_{1}  \right )^{3}}{6\left ( k_{\mathrm{GRS}}-1 \right )^{2}}.$$

By \cite{1365423} and \cite[Lemma~2.6]{Par07}, the number of monomials of degree at most $\Delta$ in $\langle\bcG\rangle$ is 
    \begin{align*}
		n\cdot \binom{m_1+2}{3}+3n\cdot \binom{m_2+2}{3} .
	\end{align*}
Therefore, 
\begin{equation*}\label{eq:IG2CNumberOfMonomialInequality}
    \begin{aligned}
		n\cdot \binom{m_1+2}{3}+3n\cdot \binom{m_2+2}{3}\geq \frac{\Delta_{2}^{3}-\left ( \Delta_{2}-\Delta_{1} \right )^{3}}{6 \left ( k_{\mathrm{GRS}}-1 \right )^{2}}.
	\end{aligned}
\end{equation*}
Let $R \coloneqq \frac{\kGRS-1}{n}$. By rearranging we obtain
\begin{equation*}
	\begin{aligned}
	\Delta_{2}^{2}-\Delta_{1}\Delta_{2}+\frac{\Delta_{1}^{2}}{3}-\frac{2n^3R^{2}\left(\binom{m_1+2}{3}+3\binom{m_2+2}{3}\right)}{\Delta_{1}}\leq 0\ .
	\end{aligned}
\end{equation*}
Solving the inequality for $\Delta_{2}$ we obtain
\begin{equation}\label{eq:IG2CDelta2UpperBound}
        \scalebox{0.85}{$\Delta_{2} \leq \frac{\Delta_{1}}{2}\left ( 1+\sqrt{-\frac{1}{3}+\frac{4}{3}\frac{n^{3}R^{2}m_{1}\left ( m_{1}+1 \right )\left ( m_{1}+2 \right )+3n^{3}R^{2}m_{2}\left ( m_{2}+1 \right )\left ( m_{2}+2 \right )}{\Delta_{1}^{3}}} \right )$.}
    \end{equation}
Note that the right-hand side of (\ref{eq:IG2CDelta2UpperBound}) is a function in $\Delta_1$ and we denote it by $\mathcal{U}\left ( \Delta_{1} \right )$. 
By taking the first derivative and the second derivative of $\mathcal{U}\left ( \Delta_{1} \right )$,  we observe that $\frac{\partial \mathcal{U}\left ( \Delta_{1} \right )}{\partial \Delta_{1}}\leq0$ for all $\Delta_1\geq m_1(\kGRS-1)$.
 Therefore, $\mathcal{U}\left ( \Delta_{1} \right )$ is a monotonically decreasing function for $\Delta_{1} \geq m_{1}\left(k_{\mathrm{GRS}}-1\right)$. Setting $\Delta_1 = m_1(\kGRS-1)$ in \eqref{eq:IG2CDelta2UpperBound}, gives an upper bound on $\Delta_2$. %
 \begin{equation*}
    	\scalebox{0.80}{$\Delta_{2} \leq \frac{m_{1}nR}{2}\left ( 1+\sqrt{-\frac{1}{3}+\frac{4}{3}\frac{n^{3}R^{2}m_{1}\left ( m_{1}+1 \right )\left ( m_{1}+2 \right )+3n^{3}R^{2}m_{2}\left ( m_{2}+1 \right )\left ( m_{2}+2 \right )}{\left ( nRm_{1} \right )^{3}}} \right )$.}
    \end{equation*}
 By Lemma~\ref{lemma:IG2CGEquiv0}, $G_{2}\left ( X,f\left ( X \right ),g\left ( X \right ) \right ) = 0$ holds when
\begin{equation}\label{eq:IG2CG2WeightedDegreeLessThanUpperBound}
    \begin{aligned}
        &m_{1}\left ( n-t \right )+m_{2}t \geq \\ 
        &\scalebox{0.9}{$\frac{m_{1}nR}{2}\left ( 1+\sqrt{-\frac{1}{3}+\frac{4}{3}\frac{n^{3}R^{2}m_{1}\left ( m_{1}+1 \right )\left ( m_{1}+2 \right )+3n^{3}R^{2}m_{2}\left ( m_{2}+1 \right )\left ( m_{2}+2 \right )}{\left ( nRm_{1} \right )^{3}}} \right )$}.
    \end{aligned}
    \end{equation}
The bound in~\eqref{eq:upperBoundOverRatioOfm2m1} is derived by solving inequality \eqref{eq:IG2CG2WeightedDegreeLessThanUpperBound} for $t$. Note that to obtain a simpler (though slightly worse) bound we omit some terms that vanish as $m_1$ increases.

\end{IEEEproof}
\subsection{Guaranteed Decoding Radius}
The bound on $t$ given in~\eqref{eq:upperBoundOverRatioOfm2m1} is derived by only considering the two polynomials of the lowest weighted degree in $\bcG$.
However, it is not guaranteed that the resultants of these two polynomials can be factorized.
The following proposition gives a condition such that there exists a pair of polynomials in $\bcG$ whose resultants are nonzero.
\begin{proposition}\label{lemma:IG2CResultant}
Consider a pair of polynomials $Q, P\in \F[X,Y,Z]$ with $Q,P\neq 0$. Let $H_{Z}\left(X,Y\right) \coloneqq \resultant\left ( Q,P;Z \right )$ and $H_{Y}\left(X,Z\right) \coloneqq \resultant\left ( Q,P;Y \right )$ (see Definition~\ref{def:Resultant}).
Then, for any $f(X),g(X)\in\F[X]$,
\begin{equation}\label{eq:IG2CResultantEq1}
	H_{Z}\left(X,Y\right)\neq 0\ \ \mathrm{and}\ \ H_{Z}\left(X,f\left ( X \right )\right)\equiv 0
\end{equation}
and
\begin{equation}\label{eq:IG2CResultantEq2}
	H_{Y}\left(X,Z\right)\neq 0\ \ \mathrm{and}\ \ H_{Y}\left(X,g\left ( X \right )\right)\equiv 0
\end{equation}
if and only if $Q\left ( X,Y,Z \right )$ and $P\left ( X,Y,Z \right )$ have no common factor which is a function in $Y$ or $Z$ and $Q\left ( X,f\left ( X \right ),g\left ( X \right ) \right )=P\left ( X,f\left ( X \right ),g\left ( X \right ) \right )=0$.
\end{proposition}
\begin{IEEEproof}
From \cite[Lemma 2.9]{Par07} (see also \cite{cox1994ideals}), the resultant $H_Z(X,Y)$ (similarly for $H_Y(X,Z$) w.r.t.~$Z$ (to $Y$) is non-zero if and only if $Q$ and $P$ do not have a common factor that is a polynomial in $Y$ (in $Z$). From Definition~\ref{def:Resultant}, we can see that a $Y$-root of $H_Z(X,Y)$ is also a $Y$-root of $Q(X,Y,Z)$ and $P(X,Y,Z)$ and vice versa. %
Therefore, $H_Z(X, f(X))\equiv 0$ if and only if $Q(X,f(X),g(X)) = P(X,f(X),g(X))=0$ and the statement of the proposition follows.
\end{IEEEproof}
We cannot guarantee that the pair of polynomials of the lowest weighted degree in $\bcG$, from which the upper bound~\eqref{eq:upperBoundOverRatioOfm2m1} is derived, do not have a common factor in $Y$ and $Z$. Therefore, the achieved decoding radius $\hat{\tau}$ of the proposed algorithm is not guaranteed to achieve this bound. However, if the decoder does not fail, the returned list $\hat{\List}$ \emph{is} guaranteed to contain all codewords within distance $\hat{\tau}$.

\begin{theorem}\label{thm:correctCWinList}
Let $\hat{\List}$ be the returned list and $\hat{\tau}$ be the achieved decoding radius of the proposed algorithm, as described in the recovery step in Section~\ref{sec:sketchAlgorithm}. Then the returned list $\hat{\List}$ contains all codewords within distance $\hat{\tau}$ from the received word.
\end{theorem}
\begin{IEEEproof}
Let $\bC$ be a codeword at distance $t < \hat{\tau}$ of the received word.
We follow the notations of Proposition~\ref{lemma:IG2CResultant}. Recall from the recovery step in Section~\ref{sec:sketchAlgorithm} that $G_r,G_s\in\bcG$ are the two polynomials which do not have a common factor in $Y$ or $Z$ and the achieved decoding radius of the proposed algorithm is
$\hat{\tau}=\frac{n-\widehat{\Delta}/{m_{1}}}{1-{m_{2}}/{m_{1}}}$. By Proposition~\ref{lemma:IG2CResultant}, $H_Z(G_r,G_s)$ and $H_Y(G_r,G_s)$ are nonzero. Then $\max\{\wdeg(G_r),\wdeg(G_s)\}=\widehat{\Delta}<m_{1}\left(n-t\right)+m_{2}t$. It follows from Lemma~\ref{lemma:IG2CGEquiv0} that $f(X)$ and $g(X)$ (the polynomials associated to the codewords $\bC$) are $Y$-root and $Z$-root of both $G_r$ and $G_s$.
By Proposition~\ref{lemma:IG2CResultant}, $f(X)$ is a $Y$-root of $H_Z(G_r,G_s)$ and $g(X)$ is a $Z$-root of $H_Y(G_r,G_s)$. Therefore, the codeword $\bC$ is in the list $\hat{\List}$.
\end{IEEEproof}

The following theorem shows that the decoding always succeeds when the number of errors is within the binary Johnson radius, which is the same radius achieved by the application~\cite{6089384} of Koetter-Vardy algorithm~\cite{algebraic03koetter} to binary Goppa codes.
\begin{theorem}\label{theorem:IG2CDecodingRadiusLowerBound}
Let $\hat{\tau}$ be the achieved decoding radius from the recovery step in Section~\ref{sec:sketchAlgorithm}. Then, for sufficiently large $m_{1}$ and $m_{2}$, we have
\begin{equation*}
	\begin{aligned}
	    \hat{\tau} > \left\lceil  \frac{1}{2}\left ( n-n\sqrt{1-\frac{2d}{n}} \right ) -1\right\rceil,
	\end{aligned}
\end{equation*}
and $\Pr\{\texttt{decoding failure}\} = 0$.
\end{theorem}
\begin{IEEEproof}
We prove this theorem by showing that the interpolation polynomials of the Koetter-Vardy algorithm are also contained in the constructed Groebner basis. The statement then follows from observing that the Koetter-Vardy algorithm has a deterministic decoding radius when $m_{1}$ and $m_{2}$ are large enough \cite{6089384}. Specifically, when $t< \frac{1}{2}\big( n-n\sqrt{1-2\frac{d}{n}} \big)$, there always exist two nonzero bivariate polynomials $U\left(X,Y\right)$ and $V\left(X,Z\right)$ of degree less than $m_{1}\left ( n-t \right )+m_{2}t$ such that $U\left(X,f\left ( X \right )\right)=0$ and $V\left(X,g\left ( X \right )\right)=0$. The polynomials $f\left ( X \right )$ and $g\left ( X \right )$ can then be obtained by factorizing $U\left(X,Y\right)$ and $V\left(X,Z\right)$.
\par
Now, we consider all $l'$ polynomials of $\boldsymbol{\mathcal{G}}$ with $\wdeg\left ( G_{i}\left ( X,Y,Z \right ) \right )<m_{1}\left ( n-t \right )+m_{2}t, \forall \ i\in\left \{ 1,2,\ldots,l' \right \}$. From Lemma \ref{lemma:IG2CGEquiv0}, we know $G_{i}\left ( X,f\left ( X \right ),g\left ( X \right ) \right )=0, \forall \ i\in\left \{ 1,2,\ldots,l' \right \}$. The proposed algorithm returns all codewords at distance $t$ to the received word if and only if these $l'$ polynomials do not have a common divisor that is a function in $Y$ or $Z$ (see the recovery step in \ref{sec:sketchAlgorithm}).

Since $\boldsymbol{\mathcal{G}}$ is a Groebner basis, $U\left(X,Y\right)$ and $V\left(X,Z\right)$ with weighted degree less than $m_{1}\left ( n-t \right )+m_{2}t$ can be represented by linear combinations of these $l'$ polynomials in $\boldsymbol{\mathcal{G}}$. We write
\begin{equation*}
	\begin{aligned}
	    &\scalebox{0.9}{$U\left(X,Y\right)=\mathrm{Coeff}_{u_{1}}G_{u_{1}}\left ( X,Y,Z \right ) + \mathrm{Coeff}_{u_{2}}G_{u_{2}}\left ( X,Y,Z \right )+\ldots$} \\
        &\scalebox{0.9}{$V\left(X,Z\right)=\mathrm{Coeff}_{v_{1}}G_{v_{1}}\left ( X,Y,Z \right ) + \mathrm{Coeff}_{v_{2}}G_{v_{2}}\left ( X,Y,Z \right )+\ldots$},
	\end{aligned}
\end{equation*}
where $u_{i},v_{j}\in\left \{ 1,2,\ldots,l' \right \}$ and the $\mathrm{Coeff}$ are some scalar coefficients.

Then, the greatest common divisor of $G_{u_{1}}\left ( X,Y,Z \right ), G_{u_{2}}\left ( X,Y,Z \right ),\ldots$ can only be a function in $X$ or $Y$ and the greatest common divisor of $G_{v_{1}}\left ( X,Y,Z \right ), G_{v_{2}}\left ( X,Y,Z \right ),\ldots$ can only be a function in $X$ or $Z$. The theorem is proved because the greatest common divisor of these $l'$ polynomials is not a function in $Y$ or $Z$. Therefore, we always obtain a non-zero resultant and the proposed algorithm returns all codewords at distance at most $t$, i.e., $\hat{\tau} > t$.

\end{IEEEproof}

\subsection{Analysis of the Size of the Returned List}
For a received word $\bR$ with $t$ errors, let $\bcG$ be the Groebner basis returned by the interpolation step and $\hat{\List}$ be the list returned by the recovery step in Section~\ref{sec:sketchAlgorithm}. We discuss the maximal list size $|\hat{\List}|$ of the proposed algorithm.
\begin{lemma}\label{lem:RootsOfResultant}
    For any pair of nonzero $Q(X,Y,Z)$, $P(X,Y,Z) \in \bcG$ and $t$ fulfilling~\eqref{eq:upperBoundOverRatioOfm2m1}, the number of $Y$-roots of $H_Z(X,Y)=\resultant(Q,P;Z)$ is upper bounded by
    \begin{equation*}
        \chi \coloneqq \left ( \frac{m_{1}\left ( n-t \right )+m_{2}t}{k_{\mathrm{GRS}}-1} \right )^{3} + 2 \left ( \frac{m_{1}\left ( n-t \right )+m_{2}t}{k_{\mathrm{GRS}}-1} \right ) .
    \end{equation*}
\end{lemma}
\begin{IEEEproof}
From Definition \ref{def:Resultant}, $H_{Z}\left(X,Y\right)$ has up to
$\mathrm{deg_{Y}}\ \left(\mathrm{Coeff}_{Q}\left(X,Y\right)\right) + \mathrm{deg_{Y}}\ \left(\mathrm{Coeff}_{P}\left(X,Y\right)\right) + \mathrm{deg_{Y}}\ \left(\prod_{i,j}\left ( q_{i}\left ( X,Y \right ) - p_{j}\left ( X,Y \right ) \right )\right)$
solutions for $Y$. We have %
\begin{equation}\label{eq:IG2CUpperBoundListSizeEq1}
	\begin{aligned}
	    \mathrm{deg_{Y}}\ \left(\mathrm{Coeff}_{Q}\left(X,Y\right)\right)&\leq \mathrm{deg_{Y}}\ \left(Q\left ( X,Y,Z \right )\right)
	    \\ &\leq\frac{m_{1}\left ( n-t \right )+m_{2}t}{k_{\mathrm{GRS}}-1}
	\end{aligned}
\end{equation}
and
\begin{equation}\label{eq:IG2CUpperBoundListSizeEq2}
	\begin{aligned}
	    \mathrm{deg_{Y}}\ \left(\mathrm{Coeff}_{P}\left(X,Y\right)\right)&\leq \mathrm{deg_{Y}}\ \left(P\left ( X,Y,Z \right )\right)
	    \\ &\leq\frac{m_{1}\left ( n-t \right )+m_{2}t}{k_{\mathrm{GRS}}-1}.
	\end{aligned}
\end{equation}
The product notation $\prod_{i,j}$ multiplies up to $\mathrm{deg_{Z}}\ \left(Q\left ( X,Y,Z \right )\right)\ \cdot \ \mathrm{deg_{Z}}\ \left(P\left ( X,Y,Z \right )\right)$ terms. For each term $q_{i}\left ( X,Y \right ) - p_{j}\left ( X,Y \right )$, there are up to $\mathrm{deg_{Y}}\ \left ( q_{i}\left ( X,Y \right ) - p_{j}\left ( X,Y \right ) \right )$ solutions for $Y$. Therefore,
\begin{equation}\label{eq:IG2CUpperBoundListSizeEq3}
	\begin{aligned}
	    \mathrm{deg_{Y}}&\ \left(\prod_{i,j}\left ( q_{i}\left ( X,Y \right ) - p_{j}\left ( X,Y \right ) \right )\right)\\
	    \leq& \ \deg_{Z}\ \left(Q\left ( X,Y,Z \right )\right)\cdot \deg_{Z}\ \left(P\left ( X,Y,Z \right )\right)\\
	    &\quad \cdot \deg_{Y}\ \left ( q_{i}\left ( X,Y \right ) - p_{j}\left ( X,Y \right ) \right )
	    \\
	    \leq &\ \left ( \frac{m_{1}\left ( n-t \right )+m_{2}t}{k_{\mathrm{GRS}}-1} \right )^{3}.
	\end{aligned}
\end{equation}
We prove the Lemma by summing up the upper bounds in (\ref{eq:IG2CUpperBoundListSizeEq1}), (\ref{eq:IG2CUpperBoundListSizeEq2}), and (\ref{eq:IG2CUpperBoundListSizeEq3}).
The Lemma is proved by studying the resultant with respect to $Z$. Note that the same result can be obtained by studying the resultant with respect to $Y$.

\end{IEEEproof}
It follows directly from the recovery step in Section~\ref{sec:algorithm} that the size $|\hat{\List}|$ of the returned list is at most $\chi$, which grows polynomially in the code parameters.

In our simulation with $n=32, d=13, t=10$, the size of the returned list is $1$ with a probability $>99.9\%$.
\subsection{Complexity Analysis}
\label{sec:complexity}
In the proposed algorithm, the interpolation step is the most computationally expensive step. In this step, we impose $n\cdot \binom{m_1+2}{3}+3n\cdot \binom{m_2+2}{3}$ linear constraints. 
The overall complexity of the interpolation step is $O\left ( n^{2}R^{-\frac{2}{3}}m_{1}^{8}+n^{2}R^{-\frac{2}{3}}m_{1}^{6}m_{2}^{2}\right )$, where $R=(\kGRS-1)/n$. 
Note that when $m_1\gg m_2$, the order of complexity of the proposed algorithm is the same as Parvaresh's algorithm for interleaved RS codes~\cite{Par07}.

\section{Choices of Multiplicities $m_1$ and $m_2$}\label{sec:choices}

The upper bound in Theorem~\ref{theorem:IG2CDecodingRadiusUpperBound} on the number of errors $t$ is a function of the ratio ${m_{2}}/{m_{1}}$. In this section, we discuss different strategies of choosing $m_1$ and $m_2$.

As described in the interpolation step in
Section~\ref{sec:sketchAlgorithm}, $m_{1}$ denotes the multiplicity assigned to the received points in $\bcP'$ and $m_{2}$ denotes the multiplicity assigned to the other points in $\cS\times \F_2\times \F_2\setminus \bcP'$. Therefore, the total number of multiplicities assigned to all points $(x,y,z)\in \cS \times \F_2\times \F_2$ is
\begin{equation*}
	\begin{aligned}
	m_{\mathrm{total}} \coloneqq m_{1}+\left ( 2^{2}-1 \right )m_{2}.
	\end{aligned}
\end{equation*}

Given a total multiplicity $m_{\mathrm{total}}$, the code length $n$, the designed minimum distance $d$, and the number of errors $t$, we discuss the following two strategies to choose the multiplicities $m_{1}$ and $m_{2}$.

\begin{figure}[t]
    \centering %
    \includegraphics[clip=true, width=\columnwidth]{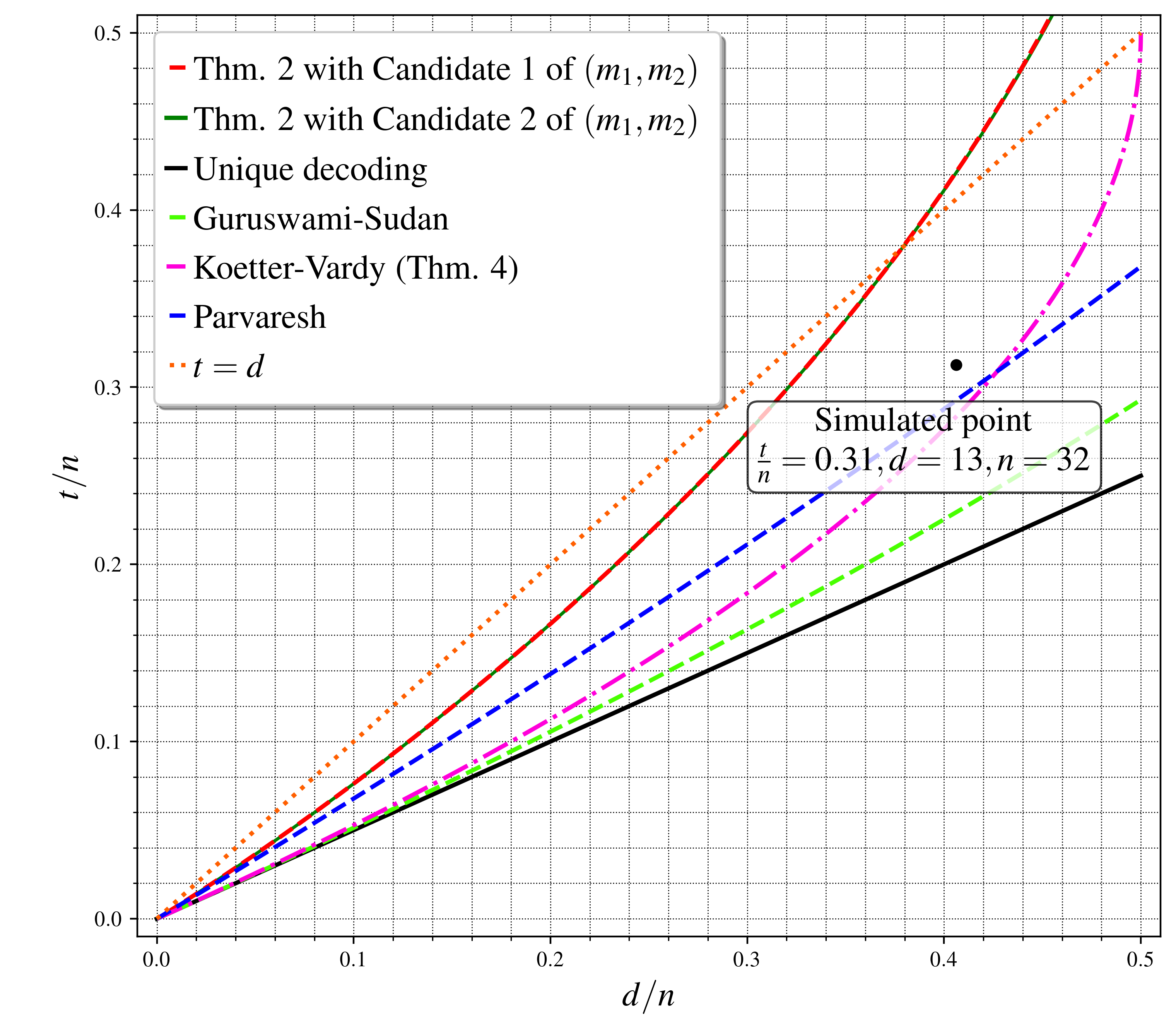}
    \caption{
    Illustrations of the upper bounds in Theorem~\ref{theorem:IG2CDecodingRadiusUpperBound} on the normalized decoding radius $t/n$ with difference candidates of $(m_1,m_2)$ and the normalized decoding radius of Guruswami-Sudan algorithm~\cite{782097} (field-size independent Johnson bound), Koetter-Vardy algorithm~\cite{algebraic03koetter} (binary Johnson bound), and Parvaresh's algorithm~\cite{Par07} (for interleaved RS codes). The $x$-axis $d/n$ is the normalized design minimum distance of the code being list-decoded. The ``Simulated point'' represents the parameters of the simulations described in Section \ref{sec:simulationResults} (see also Figure~\ref{fig:SIM32}) and illustrates that, despite the bound of Theorem~\ref{theorem:IG2CDecodingRadiusUpperBound} being an upper bound on the radius, the achieved decoding radius exceeds the radius of all known algorithms.}
    \label{fig:SIMTheory}
\end{figure}

\begin{itemize}[leftmargin=*]
    \item \emph{Candidate 1:} Choose the ratio ${m_{2}}/{m_{1}}$ such that it maximizes the upper bound~\eqref{eq:upperBoundOverRatioOfm2m1} on the decoding radius,
    i.e.,
    \begin{equation*}
    	\begin{aligned}
    	    \frac{m_2}{m_1} = \arg\max_{\frac{m_{2}}{m_{1}}\in \left [ 0,1 \right ]}\left \{ \sigma \left ( \frac{m_{2}}{m_{1}} \right ) \right \}.
    	\end{aligned}
    \end{equation*}
    with $\sigma(\frac{m_2}{m_1})$ defined as in~\eqref{eq:normalized_radius}.

    \item \emph{Candidate 2:}
    Given
    a reliability matrix $\boldsymbol{\Pi} \in\mathbb{R}^{n\times 2^{2}}$ reflecting the reliabilities of the received points,
    find a multiplicity matrix $\boldsymbol{M}\in \mathbb{N}_{0}^{n\times 2^{2}}$ as given in Algorithm~\ref{alg:IG2CCandidate3}, which is a generalization of \cite[Algorithm A]{algebraic03koetter}. By $\pi_{i,j}$ and $m_{i,j}$ we denote the entries of $\boldsymbol{\Pi}$ and $\boldsymbol{\bM}$, respectively.

    For the $2^2$-ary symmetric channel\footnote{Here, we interpret the interleaved codeword $\bC\in \F_2^{2\times n}$
    as a vector $\bc\in \F_{2^2}^n$.} with cross-over probability $t/n$, the reliability matrix $\boldsymbol{\Pi}$ is by
    \begin{equation*}
    	\begin{aligned}
    	    \pi_{i,j}= \begin{cases}
    	    1-t/n,\ &\mathrm{if}\ y_{i}\cdot 2+z_i=j,\\
            \frac{{t}/{n}}{2^{2}-1},\ &\mathrm{else}
            \end{cases},\ \forall i\in [n], j\in[4].
    	\end{aligned}
    \end{equation*}
    \par
    To find the best assignment of $m_{i,j}$ for a given number of interpolation constraints, the update of the entries $\pi_{i,j}^*$ in Line~4 should depend on the number of additional interpolation constraints imposed when updating the entry $m_{i,j}$.
    The number of interpolation constraints increases by $1/2\left ( m_{i,j}^{2}+3m_{i,j}+2 \right )$ if the entry corresponding to $m_{i,j}$ is increased to $m_{i,j}+1$. Note that the denominator of line 4 is $1/2((m_{i,j}+1)^{2}+3(m_{i,j}+1)+2)$.
    When $m_{\mathrm{total}}$ is sufficiently large, the $m_{i,j}^{2}$ term in the denominator becomes the dominant term. Since the channel is symmetric, the algorithm then returns a multiplicity matrix converging to two distinct values $m_1$ and $m_2$ with
    \begin{equation*}
     \frac{1-\frac{t}{n}}{m_{1}^{2}}\approx \frac{\frac{t/n}{2^2-1}}{m_{2}^{2}} .
    \end{equation*}
    Specifically, Algorithm~\ref{alg:IG2CCandidate3} returns $m_{1}$ and $m_{2}$ with
\begin{equation*}%
	\begin{aligned}
	    \frac{m_{2}}{m_{1}}\approx \sqrt{\frac{{t}/{n}}{3\left ( 1-{t}/{n} \right )}}.
	\end{aligned}
\end{equation*}

\end{itemize}
\begin{algorithm}[t]
\caption{Multiplicities Decision for Candidate 2}\label{alg:IG2CCandidate3}
\DontPrintSemicolon
\SetKw{Init}{Init:}
\KwIn{$\boldsymbol{\Pi}$, $m_{\mathrm{total}}, n$}
\Init{$\boldsymbol{\Pi}^{*} \gets \boldsymbol{\Pi}$, $\boldsymbol{M} \gets \boldsymbol{0}$, $s \gets m_{\mathrm{total}}\cdot n$}\\
\While{$s\neq 0$}{
Find position $\left(i,j\right)$ of the largest  entry $\pi_{i,j}^{*}$ in $\boldsymbol{\Pi}^{*}$\\
$\pi_{i,j}^{*} \gets \frac{\pi_{i,j}}{\frac{1}{2}\left(m_{i,j}^{2}+5m_{i,j}+6\right)}$\label{line:piUpdate}\\
$m_{i,j} \gets m_{i,j} + 1$\\
$s \gets s - 1$
}
\KwOut{$\boldsymbol{M}$}
\end{algorithm}
We illustrate the upper bound in Theorem~\ref{theorem:IG2CDecodingRadiusUpperBound} on the normalized decoding radius $t/n$ of the proposed algorithm with both candidates of $m_2/m_1$ in Figure~\ref{fig:SIMTheory}. For sufficiently large $m_{\mathrm{total}}$, Candidate~1 and~2 result in a similar choice for $m_1$ and $m_2$. Therefore, the upper bounds with either candidate coincide, as evident from Figure~\ref{fig:SIMTheory}, where we set $m_{\mathrm{total}}=1000$. Note that this value is chosen to illustrate the \emph{maximal} achievable radius and is generally not a practical choice for implementation.

For comparison, we also include the normalized decoding radii of other list decoding algorithms in Figure~\ref{fig:SIMTheory}. It is evident that the upper bound of the proposed algorithm is larger than the decoding radius of other existing algorithms.
Note that the Koetter-Vardy radius is also achievable by the proposed algorithm with success probability $1$ (see Theorem~\ref{theorem:IG2CDecodingRadiusLowerBound}).

\section{Simulations}\label{sec:simulationResults}
In this section, we provide simulation results for the proposed algorithm. Since the proposed algorithm does not guarantee successful decoding when $t>\frac{1}{2}\left ( n-\sqrt{n\left(n-2d\right)} \right )$, we evaluate its performance by simulating the probability of a \emph{decoding success}, which is the case that the transmitted codeword is in the returned list $\hat{\List}$.

\begin{figure}
    \centering
    \includegraphics[width=\columnwidth]{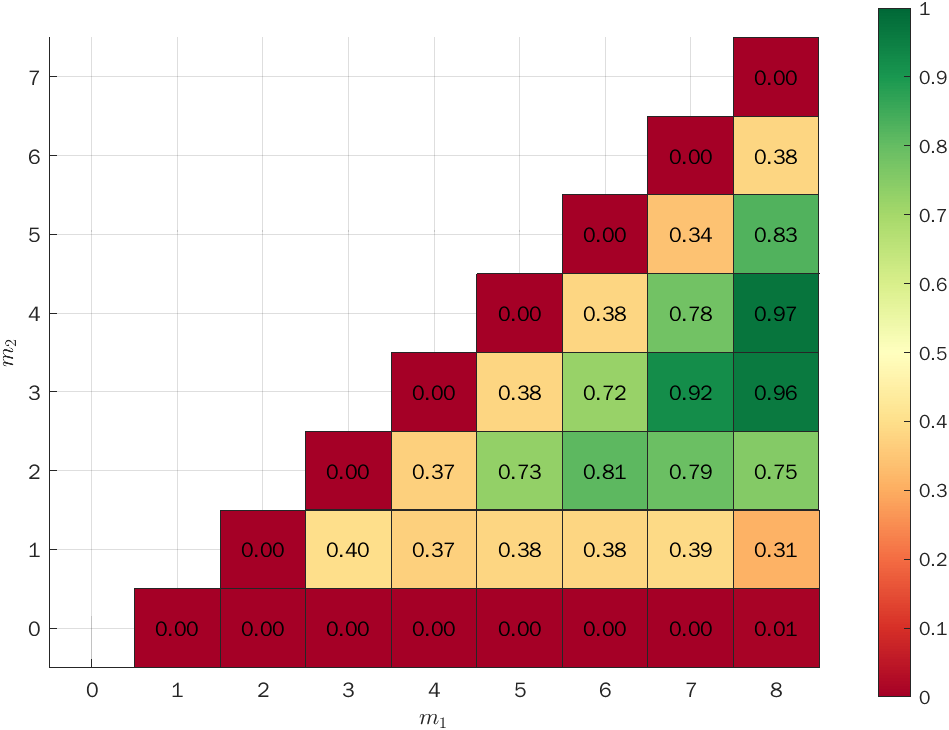}
    \caption{Probability of a decoding success for different combinations of $m_{1}$ and $m_{2}$ for $n=32, d=13$, and $t=10$. We consider a decoding success as the case that the transmitted codeword is in the list $\hat{\List}$.
    }
    \label{fig:SIM32}
\end{figure}

The black point in Figure~\ref{fig:SIMTheory} corresponds to the parameters $n=32, d=13, t=10$. As discussed in Section~\ref{sec:complexity}, $m_1$ is the dominant term in the complexity of the algorithm. Due to limitations in computational power, we chose relatively small $m_1$ and $m_2$ for the simulation and are therefore not able to achieve the upper bound given in Theorem~\ref{theorem:IG2CDecodingRadiusUpperBound}. Nevertheless, the number of errors $t=10$  is beyond the decoding radius of the existing list decoding algorithms for a binary code with $n=32$, and $d=13$.

From the results on the probability of a decoding success, illustrated in Figure~\ref{fig:SIM32}, it can be seen that, by choosing proper $(m_1,m_2)$ (e.g., $(8,3)$ or $(8,4)$), the proposed algorithm can successfully decode beyond the radius of the other list decoding algorithms with high probability. %

Moreover, we investigate the impact of $m_{1}$ and $m_{2}$ on success probability via simulations.
Figure~\ref{fig:SIM32} shows the probability of decoding success for different combinations of $m_{1}$ and $m_{2}$. We ran $\geq100$ simulations for every
pair of $(m_1,m_2)$ in Figure~\ref{fig:SIM32}.
It can be seen increasing $m_{\mathrm{total}}$ generally improves the probability of success, but only if the ratio of $m_{1}$ and $m_{2}$ is chosen suitably.

\bibliographystyle{IEEEtran}
\bibliography{6_literature}

\appendix
\section{Full Algorithms}
Here we provide the detailed algorithms that are introduced in Section~\ref{sec:sketchAlgorithm}.

\begin{algorithm}
\caption{List Decoder of $2$-Interleaved Binary Alternant Codes (Initialization)}\label{alg:IG2CInitialization}

\DontPrintSemicolon
\SetKw{Init}{Init:}
\KwIn{$\Delta$, $k_{\mathrm{GRS}}$, $\boldsymbol{\mathcal{G}}^{\left ( 0 \right )}=\left \{G_{k}^{\left ( 0 \right )}=0\right\}_{k=1}^{l}$}
\Init{$\mu \gets \left \lceil \frac{\Delta}{k_{\mathrm{GRS}}-1} \right \rceil$, $l \gets \frac{\mu \left ( \mu+1 \right )}{2}$}\\
\For {$i:=0 \ \mathrm{to}\  \mu-1$}{
    \For {$j:=0\ \mathrm{to}\ \mu-i-1$}{
         $G_{\left(i-1\right)\mu-\frac{i\left(i+1\right)}{2}+j+1}^{\left ( 0 \right )} \gets Y^{i}Z^{j}$
    }
}
\KwOut{$\boldsymbol{\mathcal{G}}^{\left ( 0 \right )}=\left \{G_{k}^{\left ( 0 \right )}\right\}_{k=1}^{l}$}
\end{algorithm}

\begin{algorithm*}
\caption{List Decoder of $2$-Interleaved Binary Alternant Codes (Interpolation)}\label{alg:IG2CInterpolation}

\DontPrintSemicolon
\SetKw{Init}{Init:}

\KwIn{$\boldsymbol{\mathcal{G}}^{\left ( 0 \right )}=\left \{G_{k}^{\left ( 0 \right )}\right\}_{k=1}^{l}$, $m_{1}$, $m_{2}$, $\boldsymbol{\mathcal{P}}$}
\Init{$i=1$}\\
\For{$\left ( x_{s},y_{s},z_{s} \right ) \  \in \  \boldsymbol{\mathcal{P}}$}{
	\For{$\left ( \gamma_{1},\gamma_{2} \right ) \  \in \  \left ( \mathbb{F}_{2}, \mathbb{F}_{2}\right )$}{
		\uIf{$\gamma_{1}=y_{s} \ \wedge\  \gamma_{2}=z_{s}$}{
			\For{$a:=0 \ \mathrm{to}\  m_{1}-1$}{
				\For{$b:=0 \ \mathrm{to}\  m_{1}-a-1$}{
					\For{$c:=0 \ \mathrm{to}\  m_{1}-a-b-1$}{
						\For{$j:=1 \ \mathrm{to}\  l$}{
						    $g_{j,p,q,r}^{\left (i-1  \right )} \gets \mathrm{coefficient\ of\ the\ term\ }X^{p}Y^{q}Z^{r}\mathrm{\ in\ }G_{j}^{\left ( i-1 \right )} \left ( X,Y,Z \right )\ \mathrm{as\ defined\ in\ Definition\ \ref{def:IG2C3DHasseDerivative}}$\\
							$\delta _{j} \gets \sum\limits_{p=a}\sum\limits_{q=b}\sum\limits_{r=c}\binom{p}{a}\binom{q}{b}\binom{r}{c}\  g_{j,p,q,r}^{\left (i-1  \right )}\  x_{s}^{p-a}{\left ( \gamma_{1}\beta_{j}^{-1} \right )}^{q-b}{\left ( \gamma_{2}\beta_{j}^{-1} \right )}^{r-c}$
						}
						$j' \gets j : \, \mathrm{The\,least\,weighted\,degree\,polynomial\,}G_{j}^{\left ( i-1 \right )} \mathrm{\,s.t.} \, \delta _{j}\neq 0 $\\
						$\mathrm{\textbf{Continue}\ if\ }\delta _{j}=0,\ \forall \ j \in \left [ 1,l \right ]$\\
						\For{$j:=1 \ \mathrm{to}\  l \  \mathrm{except} \  j'$}{
							$G_{j}^{\left ( i \right )} \left ( X,Y,Z \right ) \gets G_{j}^{\left ( i-1 \right )} \left ( X,Y,Z \right ) - \frac{\delta _{j}}{\delta _{j'}} \  G_{j'}^{\left ( i-1 \right )} \left ( X,Y,Z \right )$
						}
						$G_{j'}^{\left ( i \right )} \left ( X,Y,Z \right ) \gets \left ( X-x_{s} \right )\ G_{j'}^{\left ( i-1 \right )} \left ( X,Y,Z \right )$\\
						$i \gets i+1$
					}
				}
			}
		}
		\Else{
			\For{$a:=0 \ \mathrm{to}\  m_{2}-1$}{
				\For{$b:=0 \ \mathrm{to}\  m_{2}-a-1$}{
					\For{$c:=0 \ \mathrm{to}\  m_{2}-a-b-1$}{
						\For{$j:=1 \ \mathrm{to}\  l$}{
						    $g_{j,p,q,r}^{\left (i-1  \right )} \gets \mathrm{coefficient\ of\ the\ term\ }X^{p}Y^{q}Z^{r}\mathrm{\ in\ }G_{j}^{\left ( i-1 \right )} \left ( X,Y,Z \right )\ \mathrm{as\ defined\ in\ Definition\ \ref{def:IG2C3DHasseDerivative}}$\\
							$\delta _{j} \gets \sum\limits_{p=a}\sum\limits_{q=b}\sum\limits_{r=c}\binom{p}{a}\binom{q}{b}\binom{r}{c}\  g_{j,p,q,r}^{\left (i-1  \right )}\  x_{s}^{p-a}{\left ( \gamma_{1}\beta_{j}^{-1} \right )}^{q-b}{\left ( \gamma_{2}\beta_{j}^{-1} \right )}^{r-c}$
						}
						$j' \gets j : \, \mathrm{The\,least\,weighted\,degree\,polynomial\,}G_{j}^{\left ( i-1 \right )} \mathrm{\,s.t.} \, \delta _{j}\neq 0 $\\
						$\mathrm{\textbf{Continue}\ if\ }\delta _{j}=0,\ \forall \ j \in \left [ 1,l \right ]$\\
						\For{$j:=1 \ \mathrm{to}\  l \  \mathrm{except} \  j'$}{
							$G_{j}^{\left ( i \right )} \left ( X,Y,Z \right ) \gets G_{j}^{\left ( i-1 \right )} \left ( X,Y,Z \right ) - \frac{\delta _{j}}{\delta _{j'}} \  G_{j'}^{\left ( i-1 \right )} \left ( X,Y,Z \right )$
						}
						$G_{j'}^{\left ( i \right )} \left ( X,Y,Z \right ) \gets \left ( X-x_{s} \right )\ G_{j'}^{\left ( i-1 \right )} \left ( X,Y,Z \right )$\\
						$i \gets i+1$
					}
				}
			}
		}
	}
}
$\boldsymbol{\mathcal{G}}=\left\{G_{k}^{\left ( i \right )}\right\}_{k=1}^{l}=\left\{G_{k}\left(X,Y,Z\right)\right\}_{k=1}^{l}$\\
$\mathrm{Sort} \  \boldsymbol{\mathcal{G}} \  \mathrm{in\ the\ ascending\ order\ of\ weighted\ degree }$\\
\KwOut{ $\boldsymbol{\mathcal{G}}$}

\end{algorithm*}

\begin{algorithm*}
\caption{List Decoder of $2$-Interleaved Binary Alternant Codes (Recovery)}\label{alg:IG2CRecovery}
\DontPrintSemicolon
\SetKw{Init}{Init:}

\KwIn{ $\boldsymbol{\mathcal{G}}=\left \{G_{k}\left(X,Y,Z\right)\right\}_{k=1}^{l}$}

$Q \left ( X,Y,Z \right ) \gets  {G_{1}} \left ( X,Y,Z \right )$\\
\For{$i:=2 \ \mathrm{to}\  l$}{
	$P \left ( X,Y,Z \right ) \gets  {G_{i}} \left ( X,Y,Z \right )$\\
	$\Phi \left ( X,Y,Z \right ) \gets \mathrm{gcd}\left ( Q \left ( X,Y,Z \right ),P \left ( X,Y,Z \right ) \right )$\\
	\uIf {$\mathrm{deg_{Y}} \left( \Phi \left ( X,Y,Z \right ) \right ) = 0\  \wedge \ \mathrm{deg_{Z}}\left ( \Phi \left ( X,Y,Z \right ) \right ) = 0$}{
		$H_{Z}\left ( X,Y \right ) \gets \mathrm{Resultant}\ \left ( Q\left ( X,Y,Z \right ),P\left ( X,Y,Z \right )\ ;\ Z \right )$\\
		$\boldsymbol{\mathcal{F}_{1}} \gets \mathrm{factorize}\  H_{Z}\left ( X,Y \right ) $\\
		\For{$F_{1}\left ( X,Y \right ) \  \mathrm{in} \ \boldsymbol{\mathcal{F}_{1}}  $}{
			\If {$\mathrm{deg_{Y}}\left (  F_{1}\left ( X,Y \right ) \right ) = 1 \  \wedge \  \mathrm{deg_{X}}\left (  F_{1}\left ( X,Y \right ) \right ) < k_{\mathrm{GRS}}$}{
				$H_{Y}\left ( X,Z \right ) \gets \mathrm{Resultant}\ \left ( Q\left ( X,Y,Z \right ),P\left ( X,Y,Z \right )\ ;\ Y \right )$\\
				$\boldsymbol{\mathcal{F}_{2}} \gets \mathrm{factorize}\  H_{Y}\left ( X,Z \right ) $\\
				\For{$F_{2}\left ( X,Z \right ) \  \mathrm{in} \ \boldsymbol{\mathcal{F}_{2}} $}{
					\If {$\mathrm{deg_{Z}}\left (  F_{2}\left ( X,Z \right ) \right ) = 1 \  \wedge \  \mathrm{deg_{X}}\left (  F_{2}\left ( X,Z \right ) \right ) < k_{\mathrm{GRS}}$}{
						$f\left ( X \right ) \gets F_{1}\left ( X,Y \right ) - Y$\\
						$g\left ( X \right ) \gets F_{2}\left ( X,Z \right ) - Z$\\
						$\mathrm{Append} \  \left ( f\left ( X \right ), g\left ( X \right )  \right ) \mathrm{\ to\ the\ returned\ list\ }\hat{\List} $\\
					}
				}
			}
		}
		\textbf{break}\\
	}
	\Else{
		$U \left ( X,Y,Z \right ) \gets  \frac{Q \left ( X,Y,Z \right )}{\Phi \left ( X,Y,Z \right )}$\\
		$V \left ( X,Y,Z \right ) \gets  \frac{P \left ( X,Y,Z \right )}{\Phi \left ( X,Y,Z \right )}$\\
		$H_{Z}\left ( X,Y \right ) \gets \mathrm{Resultant}\ \left ( U\left ( X,Y,Z \right ),V\left ( X,Y,Z \right )\ ;\ Z \right )$\\
		$\boldsymbol{\mathcal{F}_{1}} \gets \mathrm{factorize}\  H_{Z}\left ( X,Y \right ) $\\
		\For{$F_{1}\left ( X,Y \right ) \  \mathrm{in} \ \boldsymbol{\mathcal{F}_{1}} $}{
			\If {$\mathrm{deg_{Y}}\left (  F_{1}\left ( X,Y \right ) \right ) = 1 \  \wedge \  \mathrm{deg_{X}}\left (  F_{1} \left ( X,Y \right ) \right ) < k_{\mathrm{GRS}}$}{
				$H_{Y}\left ( X,Z \right ) \gets \mathrm{Resultant}\ \left ( U\left ( X,Y,Z \right ),V\left ( X,Y,Z \right )\ ;\ Y \right )$\\
				$\boldsymbol{\mathcal{F}_{2}} \gets \mathrm{factorize}\  H_{Y}\left ( X,Z \right ) $\\
				\For{$F_{2}\left ( X,Z \right ) \  \mathrm{in} \ \boldsymbol{\mathcal{F}_{2}} $}{
					\If {$\mathrm{deg_{Z}}\left (  F_{2}\left ( X,Z \right ) \right ) = 1 \  \wedge \  \mathrm{deg_{X}}\left (  F_{2}\left ( X,Z \right ) \right ) < k_{\mathrm{GRS}}$}{
					    $f\left ( X \right ) \gets F_{1}\left ( X,Y \right ) - Y$\\
						$g\left ( X \right ) \gets F_{2}\left ( X,Z \right ) - Z$\\
						$\mathrm{Append} \  \left ( f\left ( X \right ), g\left ( X \right )  \right ) \mathrm{\ to\ the\ returned\ list\ }\hat{\List} $\\
					}
				}
			}
		}
	}
	$Q \left ( X,Y,Z \right ) \gets  {\Phi \left ( X,Y,Z \right )}$\\
	\If{$i==l$}{$\textbf{Return}$ \texttt{decoding failure}}
}
$\hat{\Delta} \gets \wdeg{G_{i}\left(X,Y,Z\right)}$\\
$\hat{\tau} \gets \frac{n-{\hat{\Delta}}/{m_{1}}}{1-{m_{2}}/{m_{1}}}$\\
$\textbf{Return}\ \hat{\tau}, \hat{\List}$\\
\KwOut{$\hat{\tau}, \hat{\List}$}

\end{algorithm*}

\end{document}